# MARS2020 AND MARS SAMPLE RETURN

Adrian Jon Brown
NASA Headquarters
Washington, D.C.

**Abstract**

Mars Sample Return consists of three separate missions, the first of which is the Mars2020 rover which will land at Jezero crater on February 18, 2021. We describe here our remote sensing study of a particular unit that outcrops in Jezero crater that is likely to be part of the return sample suite. We report on our efforts to characterize the olivine unit using data from the CRISM instrument, including the grain size and Fe/Mg (Fo) number of the olivine. We also discuss the astrobiological significance of the unit by analogy with the stromatolite-bearing early Archean Warrawoona group in Western Australia. We also discuss the current state of the MSR architecture.

# 1.0 Introduction

*1.1 Mars2020 Mission*

The Mars2020 rover will collect a suite of samples that will be extensively cataloged and cached for a period on the surface of Mars. These samples will eventually be returned to Earth by the Mars Sample Return mission, which is currently slated for launch in 2026. The MSR mission has consistently ranked highly in the priorities of the Planetary Science community.
We have identified a compelling rock unit associated with olivine that is the "most redshifted" (1 μm band) but that is free of alteration minerals such as clays and carbonates [1]. This new discovery was made using data from the CRISM VNIR hyperspectral instrument on MRO and shines new light on the process by which the olivine-carbonate rock [2] that has made Jezero crater the landing site of choice for the Mars 2020 rover [3].

The Martian olivine-carbonate unit is thought to be 3.82 Ga old [4], underlies the Jezero delta unit, and is also incorporated into carbonate deposits around the edge of the crater thought to be lacustrine deposits [5] or associated with a potentially volcanic unit [6] which has recently been postulated to have formed as a low temperature pyroclastic ash flow [7]. Our study has placed compositional (iron vs. how magnesium rich) and grain size constraints (relatively coarse-grained) and we will show how this unit is the same unit as the olivine-carbonate in the crater, however this unaltered olivine is not within Jezero crater itself. We therefore suggest that the newly identified subunit is a "population zero" or precursor, for more highly altered olivine seen in the crater that will be investigated by Mars2020.

This study provides further insight vis-à-vis the original chemistry of the olivine unit and what it looked like before it was altered and transported in to Jezero. Samples of this ancient olivine-carbonate unit are important to understand how habitable Mars was at this time. These rock outcrops will be investigated by Mars2020 and may eventually be returned to Earth by the Mars Sample Return mission.

In this study, we wished to examine the relationship between the spectroscopic properties of olivine throughout the Jezero watershed, to determine how these are related to clay and carbonate signatures and assess how similar the carbonates inside Jezero are to those in the watershed, and thereby provide better constraints on the origins of the olivine unit and its associated alteration mineralogy. Buffered crater counts of the fluvial valleys associated with the Jezero paleolake indicate that the Jezero hydrological system ceased activity by approximately the Noachian-Hesperian boundary, similar to the timing of other large valley network systems on Mars [8] and this may be the final time water flowed on Mars [9].

## 2.0 Methods

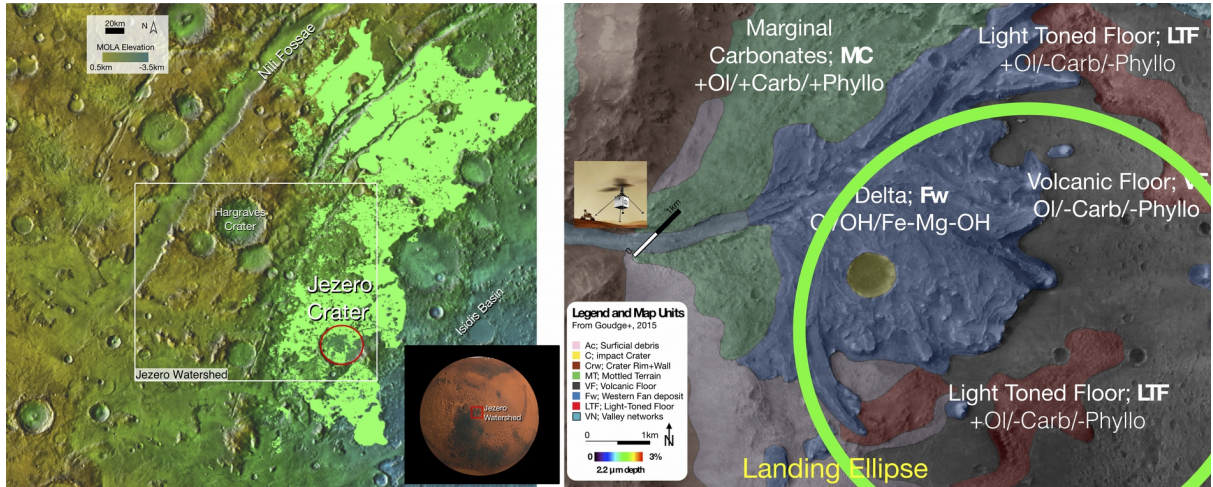

Figure 1 – (left) Study area of Jezero crater on Mars [7]. (right) The Jezero delta as mapped in Goudge et al. [6].

We used full resolution data from the CRISM instrument to map olivine and carbonate signatures present in Jezero Crater and its surrounding watershed. We used the variations in the olivine 1 μm band to place bounds on the composition and grain size and reveal a hitherto unknown relationship between the olivine 1 μm band signature and carbonates and phyllosilicates.

We used an iterative Asymmetric Gaussian approach [10-11] to track variations in band center and shape. We applied a threshold to the four parameters of the fit (centroid, asymmetry, width and amplitude) to eliminate noisy and non-olivine bearing pixels.

Edwards and Ehlmann [13] carried out a Hapke fit to a CRISM spectrum in Nili Fossae (FRT C968), north of Jezero, in the olivine-carbonate lithology. They reported an olivine of 1mm grain size and Fo60 was required to fit the olivine 1 μm band adequately The Edwards and Ehlmann spot measurement is reasonable for the location from which it was taken, however is not indicative of the full range of olivine composition. The points from 3E12 have a considerable tail that extends to the right of the [13] Fo60 estimate. This corresponds to a centroid of ~1.43μm, which intersects the red 1mm line just to the right of Fo41. Assuming a grain size of 1mm allows us to place a lower bound of Fo40 on the composition of the olivine in FRT3E12, where it is best exposed. The accuracy of our method means shows we can estimate the composition as moderate Fo (40-66).

*2.1 Correlations with carbonates and phyllosilicates*

Our study revealed the hitherto unrecognized correlations of the olivine 1 μm band position with carbonates and clays. Using "Shkuratov" [14] plots for the asymmetry and 1 μm centroid, and colorised with the 2.3 and 2.5 μm bands for our three CRISM FRT images. We showed that for the bottom row (FRT3E12), the most extreme redshifted 1 μm band pixels are not associated with carbonates or phyllosilicates (there are no red points in the blue ellipses in C and F). This behavior is not seen for HRL40FF, where the most redshifted olivines **are** accompanied by carbonates and phyllosilicates. This relationship hints at differing styles of alteration in 3E12 and

40FF, and this is likely to be investigated by the Mars 2020 rover, which will land near the location of HRL40FF.

*2.2 Links to Australian Archean Warrawoona Group*

As demonstrated in [10-12] there is a strong link between the olivine-carbonate unit at Nili Fossae and a hydrothermally altered cumulate olivine rock unit in the Archean Warrawoona Group Formation in the 3.5 Ga Pilbara region. The rocks of the Warrawoona Group constitute two komatiitic-thoelitic-felsic-chert volcanic successions which have ages spanning 3.515-3.426 Ga - as each succession gets younger in age, it gets progressively less mafic [15]. The hyperspectral VNIR signature of talc has previously been used to map a komatiite layer around the North Pole Dome in the Apex Basalt member of the Warrawoona Group [10,12]. The Apex Basalt overlies the stromatolite-bearing 3.49 Ga Dresser Formation chert-barite unit, which probably represents the late stage of an active volcanic caldera [15]. The komatiitic 3.46 Ga Apex Basalt probably represents resumption of distal volcanic activity following a ~ 20 k year hiatus. Talc-carbonate hydrothermal alteration of the Apex Basalt was either achieved on emplacement of the komatiite or when the overlying theolitic 3.46 Ga Mt. Ada Basalt unit was emplaced. Given the spectral similarities demonstrated in [10], it is likely that the Australian rocks may be of significant importance as we continue to evaluate the astrobiological potential of the Mars2020 landing site.

In situ samples of the in olivine-carbonate unit could eventually be part of the returned sample suite collected by the Mars Sample Return (MSR). We will now discuss the current structure of the Mars Sample Return mission, of which the M2020 mission is the first part.

**3.0 Mars Sample Return Architecture**

Due to the inherent complexity and expected cost of the MSR mission, this mission is international and will be carried out as a joint endeavor between NASA and the European Space Agency (ESA). NASA and ESA have already signed a Joint Statement of Intent to carry out the mission together, and are investigating the possibility of a shared sample facility that would be operated by both agencies.

NASA will be responsible for leading the campaign and providing the MSR Lander mission and construction of the Sample Capture, Handling and Containment facility within the Earth Entry vehicle. ESA would construct the Earth Return orbiter and NASA would construct the Earth Entry vehicle for the mission. ESA is also expected to construct the sample transfer arm and the "fetch rover" at the Airbus facility in Stevenage, UK, the facility that assembled the ExoMars rover. ESA has been granted initial funding to proceed with their MSR work package, and NASA has been allocated a line item in the President's budget in FY20/21 that was recently sent to Congress for consideration.

The nominal launch date is planned for 2026, with a nominal return of samples by 2031.

*3.1 MSR Rationale*

The limitations in spaceflight-ready instrumentation and the remote location of the scientific team limit the extent of scientific analyses that can be done by rover missions to Mars. This is even more likely to be the case for Mars 2020 because the sample-collecting trip involves leaving Jezero crater in a timely manner.

Inspired by the Apollo samples, which still continue propel new lunar science discoveries, we anticipate that the analyses of the samples returned by MSR will rely on future instrumentation that may not even exist today. Multiple instrument methods can be brought to bear in order to characterize the returned samples and their formation and alteration environments, which are currently poorly understood.

**4.0 Summary**

We used CRISM full resolution datasets to map variations in the shape and centroid of the olivine 1 μm band in the Jezero crater and watershed. We used the variability of the olivine 1 μm band in the Jezero watershed to place bounds on the grain size and Fo# of the olivine-carbonate lithology. Our observations show that where the olivine is best exposed (e.g. FRT 3E12) the olivine grain size must be at least 500 microns, and at most 1mm. Assuming the grain size is 1mm allows us to place a bound on the composition of the exposed olivine-carbonate lithology as Fo40-66. We have used the variations in 2.5 μm (carbonate) and 2.3 μm (clay and carbonate) bands to show that in FRT 3E12, the most redshifted olivines are not accompanied by clays or carbonates. We term these locations "population zero" olivines. This behavior is not seen in 40FF, at the location of Jezero crater.

We have discussed the connection between the Mars2020 Perseverance rover and the following elements of the Mars Sample Return mission, which will return the first intentional samples from another planet to Earth.

**Acknowledgements:** This work was supported by NAI grant #NNX15BB01A and MDAP grant #NNX16AJ48G.